\documentclass[aps,prd,floatfix,superscriptaddress,showpacs,preprintnumbers,nofootinbib,10pt]{revtex4-1}

\usepackage{epsfig}
\usepackage{dcolumn}
\usepackage{bm}
\usepackage{amsmath}
\usepackage{amsfonts}
\usepackage{amssymb}
\usepackage{subfigure}
\usepackage{graphicx}
\usepackage{latexsym}
\usepackage{slashed}
\usepackage{verbatim}
\usepackage{ulem}
\usepackage{color}
\usepackage{amsmath}
\usepackage{amsfonts}
\usepackage{amssymb}
\usepackage{graphicx}
\usepackage{caption}
\usepackage{subfigure}
\usepackage{slashed}
\usepackage{hyperref}
\def\nnd{\end{document}}

\def\be{\begin{equation}}
\def\ee{\end{equation}}

\newcommand{\bea}{\begin{eqnarray}}
\newcommand{\eea}{\end{eqnarray}}
\newcommand{\bwt}{\begin{widetext}}
\newcommand{\ewt}{\end{widetext}}

\def\u
\def\hZ{\widehat Z}

\def\eed{\end{document}}

\def\m_z{m_{\textrm {Z}}}

\renewcommand{\u}{\rm{u}}

\newcommand{\Br}{{\rm{Br}}}

\newcommand{ \slashchar }[1]{\setbox0=\hbox{$#1$}   
   \dimen0=\wd0                                     
   \setbox1=\hbox{/} \dimen1=\wd1                   
   \ifdim\dimen0>\dimen1                            
      \rlap{\hbox to \dimen0{\hfil/\hfil}}          
      #1                                            
   \else                                            
      \rlap{\hbox to \dimen1{\hfil$#1$\hfil}}       
      /                                             
   \fi}
\newcommand{\gev}{\text{GeV}}

\newcommand{\met}{\slashchar{E}_T}

\def\be{\beta}

\def\rm#1{\textrm{#1}}

\makeatother
\begin{document}

\title{The ATLAS Leptonic-$Z$ Excess from Light Squark Productions in the NMSSM Extension with a Heavy Dirac Gluino}

\author{Ran Ding}
\email[E-mail: ]{dingran@mail.nankai.edu.cn}
\affiliation{Center for High-Energy
Physics, Peking University, Beijing, 100871, P. R. China}

\author{Yizhou Fan}
\email[E-mail: ]{cosmopart2012@gmail.com}
\affiliation{State Key Laboratory of Theoretical Physics
and Kavli Institute for Theoretical Physics China (KITPC),
Institute of Theoretical Physics, Chinese Academy of Sciences,
Beijing 100190, P. R. China}

\author{Jinmian Li}
\email[E-mail: ]{jinmian.li@adelaide.edu.au}
\affiliation{ARC Centre of Excellence for Particle Physics at the Terascale (CoEPP),\\
and CSSM, Department of Physics, University of Adelaide,
South Australia 5005, Australia}

\author{Tianjun Li}
\email[E-mail: ]{tli@itp.ac.cn}
\affiliation{State Key Laboratory of Theoretical Physics
and Kavli Institute for Theoretical Physics China (KITPC),
Institute of Theoretical Physics, Chinese Academy of Sciences,
Beijing 100190, P. R. China}
\affiliation{School of Physical Electronics,
University of Electronic Science and Technology of China,
Chengdu 610054, P. R. China}

\author{Bin Zhu}
\email[E-mail: ]{zhubin@mail.nankai.edu.cn}
\affiliation{
Institute of Physics, Chinese Academy of Sciences,
Beijing 100190, P. R. China}

\begin{abstract}
  
The ATLAS Collaboration announced a 3$\sigma$ excess in the leptonic-$Z+jets+\met$ channel. We show that such an excess
can be interpreted in the extension of the Next-to-Minimal Supersymmetric Standard Model (NMSSM) with a heavy Dirac gluino
and light squarks. The abundant $Z$ bosons can be produced by light squark pair productions with the subsequent decays
$\tilde{q} \to q \tilde{\chi}_2^0 \to q Z \tilde{\chi}_1^0$.
We investigate the relevant parameter space by considering the constraints from both
the ATLAS and CMS direct SUSY searches. Our model can provide sufficient $Z$-signal events
in large parameter space if only the ATLAS searches are considered.
After combining the ATLAS and CMS searches, the maximal number of signal events can still reach about 15,
which is within the $1\sigma$ region of the observed excess. For comparison, we study the 
conventional low energy NMSSM with a Majorana gluino and its maximal number of signal events are about 11,
although we cannot realize such model in the known supersymmetry breaking scenarios.

\end{abstract}


\maketitle
\section{Introduction}
\label{introduction}

With the discovery of a 125 GeV Standard Model (SM)-like Higgs boson~\cite{Aad:2012tfa,Chatrchyan:2012ufa},
all the SM particles have already been found, and the main goal of the LHC experiment switches to search for physics
beyond the SM (BSM). As we know, supersymmetry (SUSY) is still the best-motivated framework. It provides an elegant
solution to gauge hierarchy problem, achieves gauge coupling unification at a high scale, and naturally has
the lightest supersymmetric particle (LSP) as 
a dark matter candidate if $R$ parity is preserved. However, the current null results of SUSY searches from
the LHC together with the measured 125 GeV Higgs mass have placed stringent exclusion limits on the parameter space
of its simplest realization: the Minimal Supersymmetric Standard Model (MSSM)~\cite{Martin:1997ns}. Thus,
it is worth to  consider seriously the possible extensions of MSSM, such as
the Next-to-Minimal Supersymmetric Standard Model (NMSSM) and its extensions.

Meanwhile, several recently observed excess of signal events may have already given us the new physics clues.
Notably, the ATLAS Collaboration reported a $3\sigma$ excess in the channel of two same-flavour opposite-sign
dileptons with an invariant mass around $Z$ boson $81 ~\gev < m_{\ell^+ \ell^-} < 101 ~\gev$, jets, and missing
transverse energy $\met$~\cite{Aad:2015wqa}. The goal of this search is to probe two scenarios
\begin{itemize}
\item The gluino $\tilde{g}$ pair productions and the following two-step decays through sleptons to the
  LSP neutralino $\tilde{\chi}_1^0$, which leads to the off-$Z$ dileptons.
  
\item The general gauge mediation (GGM), where gluinos first via 3-body decay into quark pairs
  and neutralino $\tilde{\chi}_1^0$, and the latter then decay into a very light gravitino $\tilde{G}$ plus a $Z$ boson,
  which results in on-$Z$ dileptons.
\end{itemize}
At the 8 TeV LHC with an integrated luminosity of $20.3 ~\rm{fb}^{-1}$, the observed number of leptonic-$Z$ events for
combined electron and muon channels  is 29 while $10.6 \pm 3.2$ events are expected in the SM~\cite{Aad:2015wqa},
which corresponds to $3\sigma$ deviation. On the other hand, the CMS Collaboration has also implemented
similar search~\cite{Khachatryan:2015lwa} and no significant signal on $Z$ excess was observed. The maximal deviation
was found $2.6 \sigma$ in the dilepton mass window $20 \gev < m_{\ell^+ \ell^-} < 70 \gev$.

Several solutions to explain this excess for both SUSY~\cite{Barenboim:2015afa,Allanach:2015xga,
  Ellwanger:2015hva,Kobakhidze:2015dra,Cao:2015ara,Cao:2015zya,Cahill-Rowley:2015cha,Liew:2015hsa,Lu:2015wwa,Collins:2015boa}
and non-SUSY models~\cite{Vignaroli:2015ama} have been proposed. Most of SUSY interpretations are based on gluino/squark
pair production processes, which can be summarized as follows
\begin{itemize}
\item The MSSM with GGM~\cite{Barenboim:2015afa}, the
 $Z$-signal can be produced via decay chain $\tilde{g}\to q \bar{q} \tilde{\chi}_1^0 \to q \bar{q} Z \tilde{G}$, where the first step decay through an off-shell squark exchange.
\item The MSSM with a light right-handed sbottom  $\tilde{b}_1$, a  bino-like LSP $\tilde{\chi}_1^0$ and
  nearly degenerated higgsino-like next-to-LSP (NLSPs) $\tilde{\chi}_{2,3}^0$~\cite{Kobakhidze:2015dra}. In this scenario, corresponding decay chain is $\tilde{g}\to b \tilde{b}_1^\dag \to b \bar{b} \tilde{\chi}_{2,3}^0 \to b \bar{b} \tilde{\chi}_{1}^0 Z$. However, such rich $b$ signals are tightly constrained by b-jet searches.
\item The MSSM with a tunned spectrum, i.e., light squraks $\tilde{q}\sim 500-750 $ GeV, bino-like NLSP $\tilde{\chi}_{3}^0\sim 350$ GeV and higgsino-like LSPs $\tilde{\chi}_{1,2}^0\sim 150-200$ GeV. The corresponding decay chain is $\tilde{q}\to q \tilde{\chi}_{3}^0 \to q \tilde{\chi}_{1,2}^0 Z$~\cite{Cahill-Rowley:2015cha}.
\item The MSSM with a split spectrum, the gluinos decay into higgsino-like NLSPs through $t$-$\tilde{t}_1$-loop, followed by higgsinos
    decay into the bino LSP plus $Z$ boson, i.e., $\tilde{g} \to g \tilde{\chi}_{2,3}^0 \to g \tilde{\chi}_{1}^0 Z$~\cite{Lu:2015wwa}.
\item The MSSM with mixed stops~\cite{Collins:2015boa}, the $Z$ boson can be produced by heavy stop decay, $\tilde{t}_2 \to Z \tilde{t}_1$, where the light stop has a mass close to the LSP in order to evade the LHC search constraints.
\item  The goldstini model with gauge mediated SUSY breaking scenario~\cite{Liew:2015hsa}, in which they also considered
  gluino radiative decay into higgsino-like neutralino $\tilde{g} \to g \tilde{\chi}_{1,2}^0$. The difference is there exists a pseudo-goldstino $G^\prime$ in the spectrum which leads to $\tilde{\chi}_{1,2}^0  \to Z G^\prime$.
\item Finally, the NMSSM with a singlino-like LSP $\tilde{\chi}_1^0$ and a bino-like NLSP $\tilde{\chi}_2^0$~\cite{Ellwanger:2015hva,Cao:2015ara,Cao:2015zya}, where extra $Z$ bosons can be produced through $\tilde{g} \to q \bar{q} \tilde{\chi}_2^0 \to q \bar{q} Z \tilde{\chi}_1^0$ or $\tilde{q} \to q \tilde{\chi}_2^0 \to q \tilde{\chi}_1^0 Z$.
\end{itemize}
In the explanation using light squarks~\cite{Cahill-Rowley:2015cha, Cao:2015zya}, the Majorana gluino mass
was chosen by hand to be heavy at low energy scale. However, we cannot realize such scenario
in the UV completion of the MSSM/NMSSM for the known supersymmetry breaking mechanisms
since the squark masses will be driven to the order of gluino mass
due to the renormalization group equation running (For an example, see Ref.~\cite{Jaeckel:2011wp}.).
In this work, we investigate an extension of the NMSSM with a Dirac gluino, which provides a UV completion
of the above scenario. In particular, the heavy Dirac gluino will not induce the SUSY electroweak
fine-tuning problem, and then such extension is still natural~\cite{Fayet:1978qc,Polchinski:1982an,Hall:1990hq, Jack:1999ud,Jack:1999fa,Fox:2002bu,Carpenter:2005tz,Hisano:2006mv,Hsieh:2007wq,Blechman:2008gu,Choi:2008pi,Amigo:2008rc,
Benakli:2008pg,Choi:2008ub,Kribs:2009zy,Belanger:2009wf,Benakli:2009mk,Chun:2009zx,Benakli:2010gi,Choi:2010gc,
Carpenter:2010as,Kribs:2010md,Abel:2011dc,Benakli:2011kz,Heikinheimo:2011fk,Kribs:2012gx,Goodsell:2012fm,
Fok:2012fb,Benakli:2012cy,Abel:2013kha,Kribs:2013eua,Kribs:2013oda,Buckley:2013sca,Dudas:2013gga,Bertuzzo:2014bwa,Benakli:2014cia, Goodsell:2014dia,Busbridge:2014sha,Nelson:2015cea,Carpenter:2015mna,Ding:2015wma}.
We demonstrate the leptonic-$Z$ excess can be successfully explained through light squark pair productions.

Before studying the model in details, we would like to discuss the decay patterns of the
first two generation squarks in the context of the MSSM and explain our strategy to address the $Z$ excess.
With heavy gluino, squarks decay into quarks plus neutralino/chargino: $\tilde{q} \to q \tilde{\chi}_i^0$
or $\tilde{q} \to q^\prime \tilde{\chi}_i^{\pm}$. The decay channel $\tilde{q} \to q \tilde{\chi}_1^0$
is always kinematically favored and for right-handed squarks it can be dominant
since in most case $\tilde{\chi}_1^0$ is bino-like. However, the left-handed squarks strongly prefer
to decay into the wino-like charginos/neutralinos due to large squark-quark-wino couplings. Moreover,
squarks decay into higgsino-like charginos/neutralinos are only important for the third generation squarks
in which squark-quark-higgsino couplings are sizeable because of large
Yukawa couplings. Due to these properties, when the LSP neutralino is bino-like, the NLSP neutralino
is wino-like (higgsino-like), squarks can not have large branching ratio for cascade decay
$\tilde{q}\to q \tilde{\chi}_2^0 (\tilde{\chi}_{2,3}^0) \to q \tilde{\chi}_1^0 Z$. While in case of
the bino-like NLSP and higgsino-like LSP, one may potentially has considerable branching ratio
for above cascade decays but the specific fine-tuning of the mass spectrum is required. Unfortunately,
above situation is also true for the MSSM with a Dirac gluino. So we pay our attention to the NMSSM with
the Dirac gluino. In this case the LSP (NLSP) neutralino is singlino-like (bino-like) and
both $\tilde{q}\to q \tilde{\chi}_2^0$ and $\tilde{\chi}_2^0 \to \tilde{\chi}_1^0 Z$ can have
large branching ratios in large parameter space.

This work is organized as follows. In Section~\ref{model}, we present our model and demonstrate its key features.
In Section~\ref{signal}, we describe the cutflow of the ATLAS search for the leptonic-$Z+jets+\met$ signal
and give the corresponding best fit benchmark points. We scan the relevant SUSY parameter space to study
the capability of our model in interpreting the ATLAS $Z$-peaked excess by considering the constraints from both
the ATLAS and CMS SUSY searches. As a comparison, we also discuss the same signal channel in the usual NMSSM
within the same parameter space. Finally, our conclusion is given in Section~\ref{conclusion}.

\section{The NMSSM Extension with a Heavy Dirac Gluino and Light Squarks}
\label{model}

Dirac gauginos have been proposed decades ago~\cite{Fayet:1978qc,Polchinski:1982an,Hall:1990hq}, and recently
have been studied extensively in model building and phenomenology~\cite{Jack:1999ud,Jack:1999fa,Fox:2002bu,Carpenter:2005tz,Hisano:2006mv,Hsieh:2007wq,Blechman:2008gu,Choi:2008pi,Amigo:2008rc,
Benakli:2008pg,Choi:2008ub,Kribs:2009zy,Belanger:2009wf,Benakli:2009mk,Chun:2009zx,Benakli:2010gi,Choi:2010gc,
Carpenter:2010as,Kribs:2010md,Abel:2011dc,Benakli:2011kz,Heikinheimo:2011fk,Kribs:2012gx,Goodsell:2012fm,
Fok:2012fb,Benakli:2012cy,Abel:2013kha,Kribs:2013eua,Kribs:2013oda,Buckley:2013sca,Dudas:2013gga,Bertuzzo:2014bwa,Benakli:2014cia,
Goodsell:2014dia,Busbridge:2014sha,Nelson:2015cea,Carpenter:2015mna,Ding:2015wma}.
The SUSY models with Dirac gluinos have notable advantages comparing with their Majorana counterparts.
Firstly, they have special renormalization properties, the so-called supersoft behavior~\cite{Jack:1999ud,Jack:1999fa,Fox:2002bu,Benakli:2011kz,Goodsell:2012fm} which allows specific mass spectra where squarks are much lighter than glunio.  Secondly, some subprocesses involved in the first two generations of squarks pair productions vanish
(those final states with squarks of the same chirality) since the $t$-channel Dirac gluino exchange
cannot mediate the chirality-flip processes.
This leads to the reduction of squark pair production cross sections in Dirac case compared to the Majorana case,
which will significantly alleviate their bounds from the current LHC direct SUSY searches, especially in
the heavier Dirac gluino region ($M_{\tilde{g}}\gtrsim 2-3$ TeV)~\cite{Heikinheimo:2011fk,Kribs:2012gx,Kribs:2013eua,Kribs:2013oda}. This is known as supersafe behavior and play a crucial role when using our model to explain
the ATLAS $Z$-peaked excess. In terms of supersafe, the first and second generations of squarks can be as light as 700 GeV, which actually removes the constraints as we want. Therefore, the appropriate and convenient framework for weak scale SUSY is the unification of Dirac gluinos and semi-soft SUSY breaking. In order to generate the Dirac gluino mass, we need to introduce a chiral superfield $\Phi $ with  $SU(3)_C\times SU(2)_L\times U(1)_Y$ quantum number
$(\mathbf{8}, \mathbf{1}, \mathbf{0})$.

In this paper, we consider the general NMSSM framework whose superpotential is given by~\cite{Ellwanger:2009dp}
\begin{align}
W&= Y_u Q H_u \bar u - Y_d Q H_d \bar d - Y_e  L H_d \bar e + (\mu + \lambda S)H_u \cdot H_d + \xi_F S + \frac{1}{2}\mu^{\prime} S^2 + \frac{1}{3} \kappa S^3,
\end{align}
where $Q,~\bar u, ~\bar d, ~L,~\bar e$, and $S$ are the NMSSM chiral superfields, $\lambda$ and $\kappa$
are the dimensionless Yukawa couplings, $\mu$ and $\mu^{\prime}$ are the bilinear mass terms, and $\xi_F$
is the tadpole term. The corresponding SUSY breaking soft terms are
\begin{align}
-\mathcal{L}_{soft}&=T_u Q H_u \bar u - T_d Q H_d \bar d - T_e L H_d \bar e + T_{\lambda} H_u \cdot H_d S +\frac{1}{3}T_{\kappa}S^{3}+ B_\mu H_u \cdot H_d +\frac{1}{2}m^{\prime2}_{S} S^2+ \xi_S S + h.c.  \nonumber\\
&+m_{H_u}^2 |H_u|^2+m_{H_d}^2 |H_d|^2+m_S^2 |S|^2+ m^2_Q|Q^2|+m^2_L|L^2|+m^2_{\bar u}|{\bar u}^2|+m^2_{\bar d}|{\bar d}^2|+m^2_{\bar e}|{\bar e}^2|\nonumber\\
&+ M_1 B B + M_2 W W + M_3 G G +M^B_{D} B \tilde{S} + M^G_{D} G \Phi +h.c. ,
\label{lagrangian}
\end{align}
where $T_{u,d,e,\lambda,\kappa}$ are the trilinear soft terms, $B_{\mu}, m^{\prime2}_{S}$ are the bilinear soft terms,
 $M_{1,2,3}$ are respectively the soft Majorana gaugino masses for bino, wino and gluino, as well as
the Dirac type of gluino (bino) mass $M^G_{D}$ ($M^B_{D}$). Noticed that in Eq.~\ref{lagrangian},
one could has mixed mass gluino eigenstates, i.e., pseudo-Dirac gluino, while
the pure Dirac  or Majorana gluino corresponds to $M_3=0$ or $M^G_{D}=0$. For simplicity, we consider
the pure Dirac gluino and set $M_3=M^B_{D}=0$.

\begin{figure} [htbp]
\begin{center}
\includegraphics[width=0.60\linewidth]{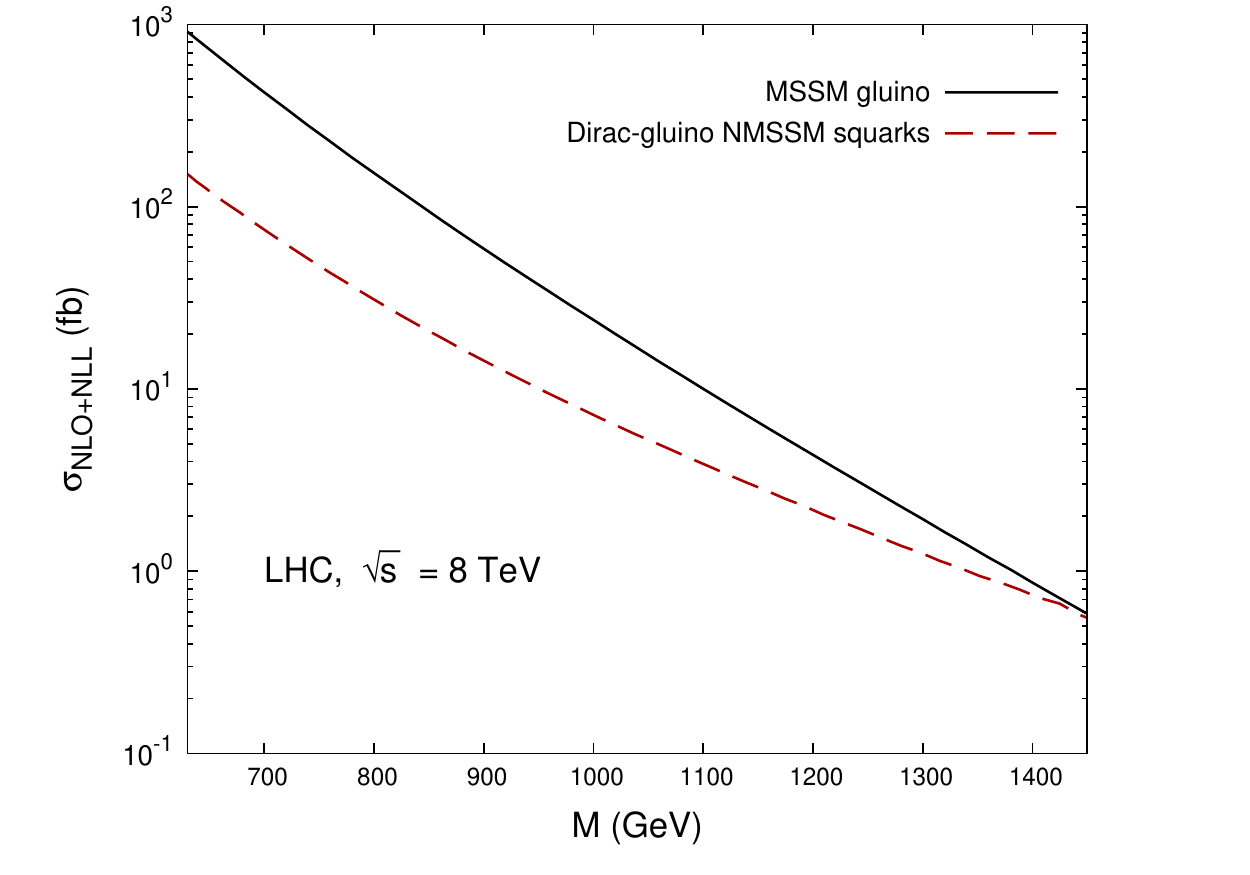}
\end{center}
\caption{The total NLO+NLL cross section of squark pair productions $\sigma_{\rm NLO+NLL}=\sigma(\tilde{q}\tilde{q}+\tilde{q}\tilde{q}^*+\tilde{q}^*\tilde{q}^*)$ in our model at the 8 TeV LHC (red dashed line). For comparison, the NLO+NLL
  cross section of gluino pair production for decoupled squarks is also shown (black solid line).
\label{sigma}}
\end{figure}

In Fig.~\ref{sigma}, we show the total cross sections of the first two generation squark pair productions
($\tilde{q}\tilde{q}+\tilde{q}\tilde{q}^*+\tilde{q}^*\tilde{q}^*$) at the
8 TeV LHC. For comparison, the NLO+NLL cross section of Majorana gluino pair production based
on {\tt NLL-fast}~\cite{NLL-fast} calculation is also presented. We find that the cross sections of
total squark pair productions are larger than 10 fb for $m_{\tilde{q}}\lesssim 900$ GeV.
Without taking into account cut efficiency and branching ratio suppression in each step of squark cascade
decay chain, we may obtain sufficient $Z$ events in this mass region. The crucial point is then
how to keep the $Z$ yield in squark cascade decays and escape the constraints from the ATLAS and CMS
direct SUSY searches. To achieve this goal, we consider the following strategy
\begin{itemize}
\item The neutralino sector is similar to that of Refs.~\cite{Ellwanger:2015hva,Cao:2015ara}, i.e.,
  the bino-like NLSP $\tilde{\chi}^0_2$ and singlino-like LSP $\tilde{\chi}^0_1$. By choosing $M_2 \sim \mu \gg M_1$, the branching ratios of squark decays into wino and Higgsino are suppressed. As a consequence, the branching ratio of
  squark decay into bino is almost $100\%$.

\item For simplicity, we assume slepton $\tilde{\ell}$ and the third generation squarks
  $\tilde{t}_{1,2}/\tilde{b}_{1,2}$ are decoupled. Also, to keep the nice feature
  of heavy Dirac gluino model, gluino mass is fixed at $m_{\tilde{g}}\gtrsim 2.6$ TeV.

\item The mass splitting between $\tilde{\chi}_2^0$ and $\tilde{\chi}_1^0$ is taken to be
  $m_{\tilde{\chi}^0_2} - m_{\tilde{\chi}_1^0}\simeq 100$ GeV to forbid $\tilde{\chi}_2^0 \to \tilde{\chi}_1^0 h$ decay,
  which guarantees $ \Br(\tilde{\chi}_2^0 \to Z \tilde{\chi}_1^0)\sim 100\%$~\footnote{This can also be realized by choosing specific SUSY parameters such that the coupling $h \bar{\tilde{\chi}}_1^0 \tilde{\chi}_2^0$ is suppressed.}.
\end{itemize}
In the next section, we present our simulation results for both $Z$ excess signal events and LHC constraints.

\section{The $Z$-Peaked Excess and LHC SUSY Search Constraints}
\label{signal}

For the purpose of studying collider phenomenology, we implement the model in the Mathematica package {\tt SARAH}\cite{Sarah},
which generates {\tt SPheno}\cite{SPheno} output file to calculate the particle mass spectra, mixing matrices
as well as the low energy constraints. It can also create the corresponding {\tt UFO}\cite{Degrande:2011ua} model file
which can be used by {\tt MG5\_aMC}~\cite{Mg5}. In our simulation, the parton level events are generated by {\tt MG5\_aMC},
whose output {\tt LHE} file then feeds into {\tt Pythia8}~\cite{Sjostrand:2007gs} to implement particle decays,
parton showering and hadronization~\footnote{In order to employ {\tt Pythia8} to correctly handle the
  decays of all SUSY particles in our model, we use the so-called ``semi-internal processes" which was
  introduced in Ref.~\cite{Desai:2011su}.}, and the package {\tt NLL-fast} is used to calculate
the NLO+NLL $k$-factor of the squark pair productions. Moreover, we use {\tt Delphes3} \cite{Delphes} for fast detector simulation.

The ATLAS and CMS Collaborations have carried out copious SUSY searches at the LHC. The null signals for most of the searches
so far have put quite stringent bounds on many SUSY models. As a result, those scenarios, which could potentially explain
the $Z$ boson excess, may have already been excluded. In order to use the LHC SUSY searches to constrain
our interesting models, we use a private package in which most of the current LHC SUSY searches have been recasted.
The package was first used in Ref.~\cite{Cheng:2013fma} and further developed in Refs.~\cite{Guo:2013asa,Cheng:2014taa}.
The check of validation can be found therein as well.

Specifying to our current work, we find that the LHC searches for the final states with jets and $\met$ and
with~\cite{Aad:2015mia,Khachatryan:2015lwa} and without~\cite{Aad:2014wea,Aad:2013wta} isolated leptons are related to 
the searches for the final states with on-shell $Z$ boson~\cite{atlas:z}. We recast those searches by strictly
following the analyses in the corresponding conference notes or published papers. The Delphes with default ATLAS setup
is used for fast detector simulation.

In order to measure whether a model is excluded by the current LHC searches or not, we define variables
$R^i \equiv \frac{N_{\text{NP}}^i}{N^i_{\text{UP}}}$, where the $N_{\text{NP}}^i$ and $N^i_{\text{UP}}$ are the numbers
of new physics event and upper limit in each signal region $S_i$ for a given analysis, respectively.
The largest $R^{\max} = \max_{i} (R^i)$ among all signal regions  of all relevant searches is specified to
the corresponding model. As a result, $R^{\max}>1$ means that there is at least one signal region in which
the number of signal events is larger than the upper limit of current data, so this model is excluded.
In the LHC  analyses, the ATLAS Collaboration has explicitly given $N^i_{\text{UP}}$ for each signal region,
while the CMS Collaboration only presented the observed numbers of events and corresponding background events
and its uncertainty.
We calculate the 95\% $C.L.$ $N^i_{\text{UP}}$ for the CMS analyses via standard Bayesian procedure as follows
\begin{align}
\frac{1}{\mathcal{N}} \int^{N_{\text{UP}}}_{0} \mathcal{L}(n_{obs}| N_{s}, N_b, \sigma_b) P(N_{s}) dN_s=0.95, \label{bayes}
\end{align}
where $P(N_{s})$ is uniform prior probability  and  $\mathcal{N}=\int^{\infty}_{0} \mathcal{L}(n_{obs}| N_{s}, N_b, \sigma_b) P(N_{s}) dN_s$ is a normalization factor. Assuming Gaussian distribution for background and signal uncertainties, the likelihood function in Eq.~\ref{bayes} can be written as
\begin{align}\label{ba}
\mathcal{L}(n_{obs}| N_{s}, N_b, \sigma_b) = \frac{1}{\sqrt{2 \pi \sigma_s^2} \sqrt{2 \pi \sigma_b^2}} \int^{5 \sigma_s}_{-5 \sigma_s} ds \int^{5 \sigma_b}_{-5 \sigma_b} db P(n_{obs};\mu) e^{\frac{db^2}{2 \sigma_b^2}} e^{\frac{ds^2}{2 \sigma_s^2}} ~.~\,
\end{align}
Practically, the probability function $P(n_{obs};\mu)$ is taken as  Possion distribution ${\mu^{n_{obs}} e^{- \mu}}/{n_{obs}!}$ for  $n_{obs} \leqslant 100$ while taken as Gaussian distribution ${e^{(n_{obs}-\mu)^2/2\mu}}/{\sqrt{2 \pi \mu}}$ for  $n_{obs} >100$. Here, the expectation value $\mu= N_s + ds + N_b + db$, and the signal error  $\sigma_s=0.01 \ n_s$. We find the derived $N^i_{\text{UP}}$ from Eq.~(\ref{bayes}) will not change much as long as the signal uncertainty is within a few tens of percent. The calculated $N^i_{\text{UP}}$s for all signal regions in Ref.~\cite{Khachatryan:2015lwa} are shown in Table~\ref{nup} for illustration.

\begin{table}[htbp]
\caption{\label{nup} The 95\% $C.L.$ $N^i_{\text{UP}}$ for all signal regions in the CMS search for events with two leptons, jets, and missing transverse momentum~\cite{Khachatryan:2015lwa}.  }
\begin{tabular}{|c|c|c|c|c|c|c|}
\hline \hline
 $N_{\text{jets}}$ & \multicolumn{3}{|c|}{$\geq 2$} & \multicolumn{3}{|c|}{$\geq 3$}  \\ \hline
 $E_T^{\text{miss}}$ (GeV) & $100-200$ & $200-300$ & $>300$ & $100-200$ & $200-300$ & $>300$ \\ \hline
 Total background & $1204 \pm 106$ & $74.5 \pm 11.3$ & $12.8 \pm 4.3$ & $478 \pm 43$ & $39.2 \pm 6.6$ & $5.3 \pm 2.3$ \\ \hline
 Data & 1187 & 65 & 7 & 490 & 35 & 6 \\ \hline
 $N_{\text{UP}}$ & 211 & 23.2 & 8.6 & 107 & 16.5 & 8.9 \\ \hline \hline
\end{tabular}
\end{table}

With these necessary tools, we are able to apply relevant LHC SUSY search constraints on all benchmark models.
We now briefly describe the cuts flow of the ATLAS analysis on $Z$ excess. According to the search,
the signal events should satisfy the following requirements
\begin{enumerate}
  \item {Events contain at least two same-flavour opposite-sign leptons.}
  \item {The leading lepton $p^{\ell,1}_{T} > 25$ GeV and sub-leading lepton $p^{\ell,2}_{T} > 10$ GeV.}
  \item {The invariant mass of these two leptons must fall into $Z$ boson mass window $81
< m_{\ell^+\ell^-} < 101$ GeV.}
  \item {Events further have $\geq 2$ jets with $p^j_{T} > 35$ GeV and $|\eta|<2.5$.}
  \item {The missing transverse energy $\met>225$ GeV.}
  \item {$H_{T}>600$ GeV with $H_{T}\equiv p^{\ell,1}_{T}+p^{\ell,2}_{T}+\sum_i p^{j,i}_{T}$.}
  \item {$\Delta\phi(j_{1,2},\met)>0.4$, where $\Delta\phi$ is the azimuthal angle between two objects in parenthesis.}
\end{enumerate}

\begin{figure} [htbp]
\begin{center}
\includegraphics[width=0.40\linewidth]{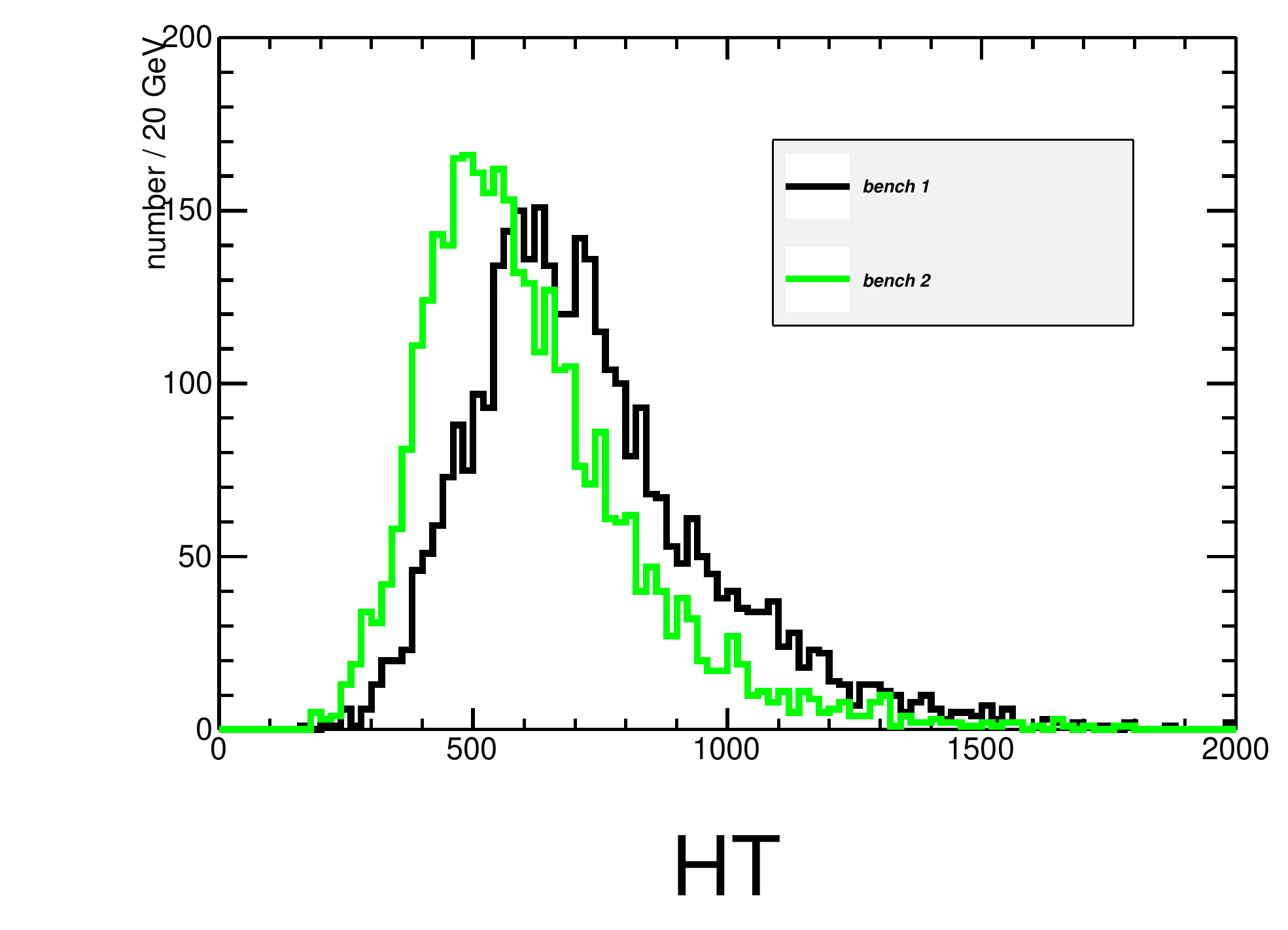}
\includegraphics[width=0.40\linewidth]{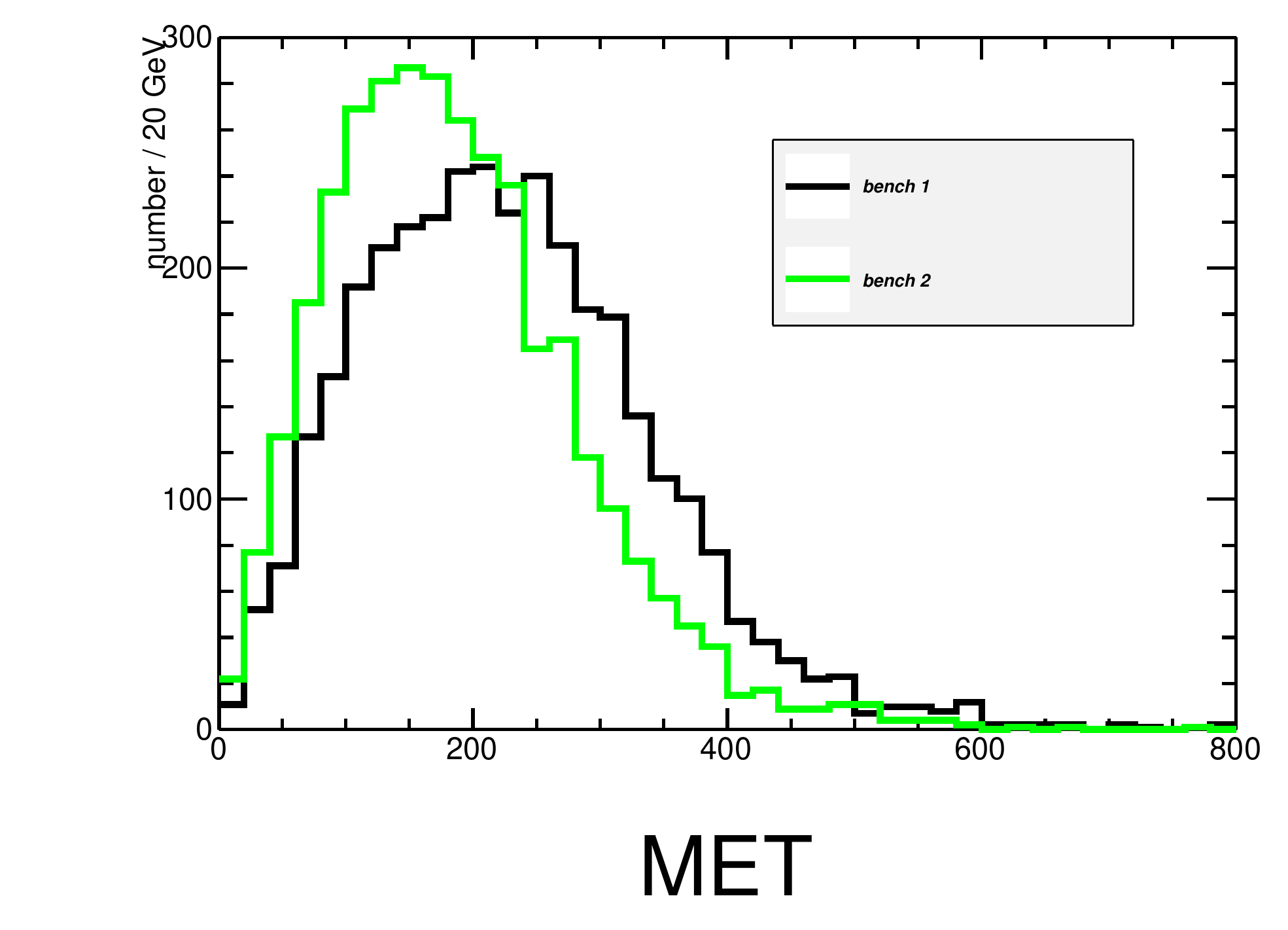}
\end{center}
\caption{$H_T$ (left) and $\met$ (right) distributions after implementing the ATLAS cut flow, except
  the cuts on $H_T$ and $\met$. The \textbf{bench1} and \textbf{bench2} are plotted by black  and green solid lines,
  respectively.
\label{htmet}}
\end{figure}

In Table~\ref{bench}, we present the most relevant information for  two  benchmark points. Their cut efficiency,
survival number of signal events in the ATLAS cut flow as well as the corresponding $R^{\max}$ values
are shown in Table~\ref{cutflow}. From this table, we find that \textbf{bench1} can give the best fit signal events (18.7)
and satisfy all of the ATLAS constraints we have considered, but not the CMS constraints.
While \textbf{bench2} can escape the constraints form both
the ATLAS and CMS searches. As the price, the signal events drop to 14.2, which roughly corresponds to $2\sigma$ deviation
from the SM and is within $1\sigma$ region of the observed excess. Finally, Fig.~\ref{htmet} shows the $H_T$ and $\met$
distributions respectively for two benchmark points after applying all cut selections
in the ATLAS cut flow except the $H_T$ and $\met$ cuts. Some important features can be learnt from here
\begin{itemize}
\item From Table~\ref{cutflow}, we find that $H_T$ and $\met$ cuts are crucial in the cut flow,
  and hard $H_T$ and $\met$ distributions are required in order to keep sufficient survival events.
\item As shown from Table~\ref{bench} and Fig.~\ref{htmet}, \textbf{bench2} has relatively
  compressed spectrum which leads to softer $H_T$ and $\met$ distributions compared with \textbf{bench1}.
\end{itemize}
As a consequence, \textbf{bench2} is easier to evade the constraints from the LHC searches especially
when the CMS limits are taken into account.
Meanwhile, fewer events  can pass $H_T$ and $\met$ cuts in the ATLAS leptonic-$Z+jets+\met$ search.
On the other hand, \textbf{bench1} easily satisfies the requirement of $Z$-signal events but is challenged by
the CMS constraints.

\begin{table}[htbp]
\caption{The relevant particle masses, branching ratios, and NLO+NLL cross sections for
two benchmark points of our model.}
\begin{center}
\begin{tabular}{|c||c|c|c|c|c|c|c|}
\hline
&$M_{\tilde{q}^{1,2}_{L,R}}$ & $M_{\tilde{\chi}^0_{1}}$ & $M_{\tilde{\chi}^0_{2}}$ & $M_{h}$ & Br(${\tilde{q}_{L,R}\to q \tilde{\chi}^0_{2}}$) & Br(${\tilde{\chi}^0_2 \to Z  \tilde{\chi}^0_{1}}$) & $\sigma_{\rm{NLO+NLL}}$ (fb)\\
\hline
\textbf{bench1}&$693$ &$283$ &$398$ &$125$ & $100\%$ &$95.7\%$ &$79$\\\hline
\textbf{bench2}&$642$ &$313$ &$428$ &$125$ & $100\%$ &$95.6\%$ &$133$\\\hline
\end{tabular}
\label{bench}
\end{center}
\end{table}

\begin{table}[htbp]
  \caption{The cut efficiency and signal events for our model in the ATLAS cut flow at the 8 TeV LHC
    with luminosity ${\cal L}=20.3$ fb$^{-1}$. The $R^{\max}$ values are also presented for
    the ATLAS searches alone as well as for both the ATLAS and CMS searches.}
\begin{center}
\begin{tabular}{|c||c|c|c|c|c|c|c|c||c|c|c|c|}
\hline
 & All Events& Cut1 & Cut2 & Cut3 & Cut4 & Cut5 & Cut6 & Cut7 & $N_{\rm{ Sig}}$ & $R^{\max}_{\rm{ATLAS}}$ & $R^{\max}_{\rm{ATLAS+CMS}}$  \\
\hline
\textbf{bench1}&$10^5$ &$3869$ &$3839$ &$3549$ &$3482$ &$1599$ &$1252$ &$1165$ & $18.7$ &$0.56$ & $1.26$  \\\hline
\textbf{bench2}&$10^5$ &$3851$ &$3822$ &$3530$ &$3441$ &$996$ &$574$ &$528$ &$14.2$ &$0.52$  &$0.96$  \\\hline
\end{tabular}
\label{cutflow}
\end{center}
\end{table}

\begin{figure} [htbp]
\begin{center}
\includegraphics[width=0.60\linewidth]{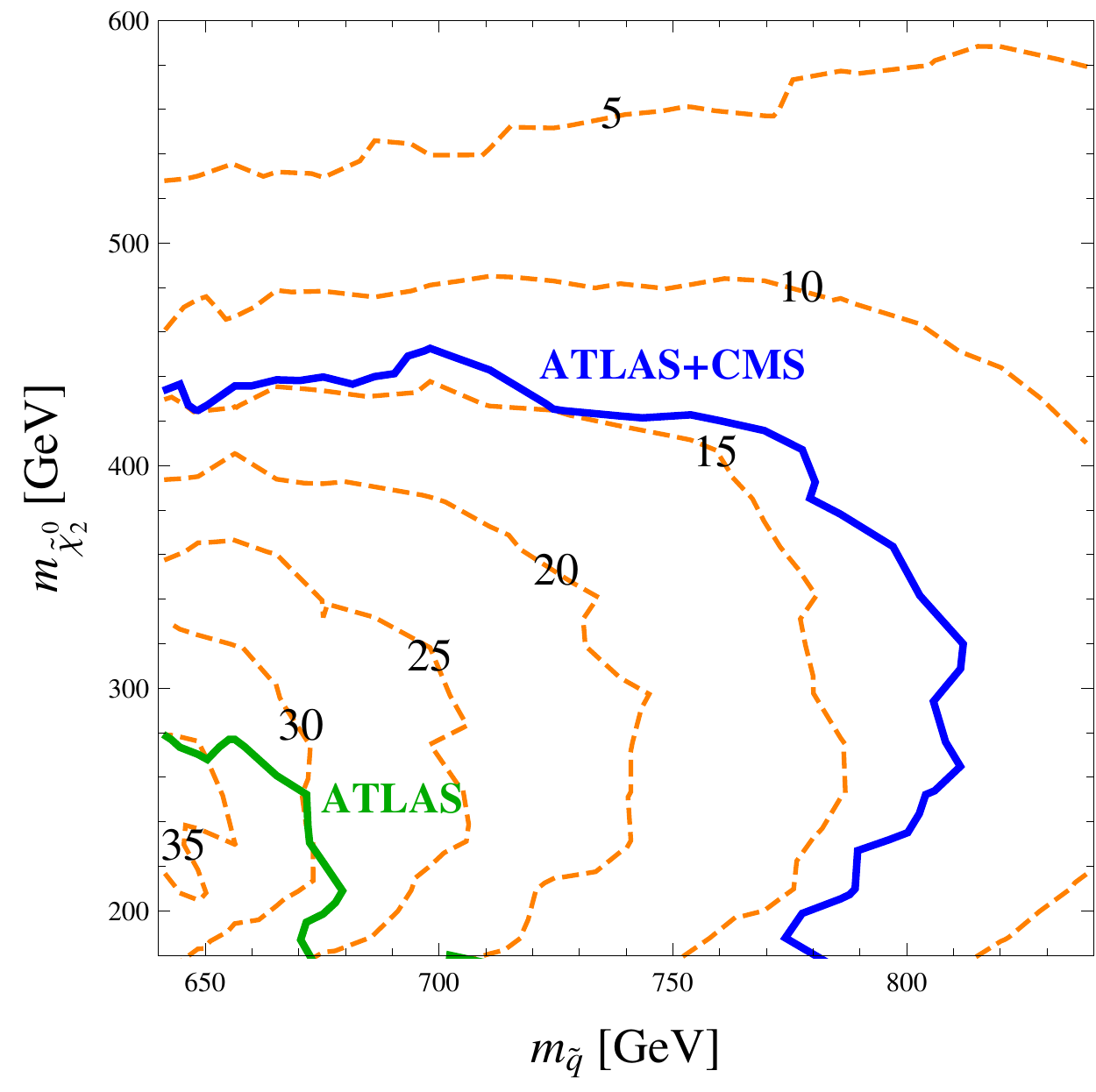}
\end{center}
\caption{Contours of signal events for the ATLAS leptonic-$Z+jets+\met$ search in the $m_{\tilde{q}} - m_{\chi^0_2}$ plane
  (orange dashed curves) for our model. The corresponding exclusion limits for the ATLAS constraints alone (green curve)
  and the  ATLAS+CMS constraints  (blue curve) are  shown as well.
\label{fig:scan_Di}}
\end{figure}

We further implement a grid scan in the $m_{\tilde{q}} - m_{\chi^0_2}$ plane. Contours of signal events and corresponding exclusion limits
are shown in Fig.~\ref{fig:scan_Di}. Among all searches undertaken in this study, we find that the CMS search for final states
with jets, $\met$ and two opposite-sign same-flavor leptons at the $Z$ pole~\cite{Khachatryan:2015lwa} gives the most stringent
bound on most of the models that could explain the ATLAS excess because of the similar signature they are looking for.
The exclusion curves in Fig.~\ref{fig:scan_Di} are shown with (blue curve marked as ATLAS+CMS) and without
(green curve marked as ATLAS) the CMS constraint, respectively. From the figure, we conclude that after taking into
account all the current LHC SUSY searches, our model can give at most $\sim 15$ events for the ATLAS $Z$-peaked excess,
which means the excess can be addressed within 1$\sigma$ level.

\begin{figure} [htbp]
\begin{center}
\includegraphics[width=0.60\linewidth]{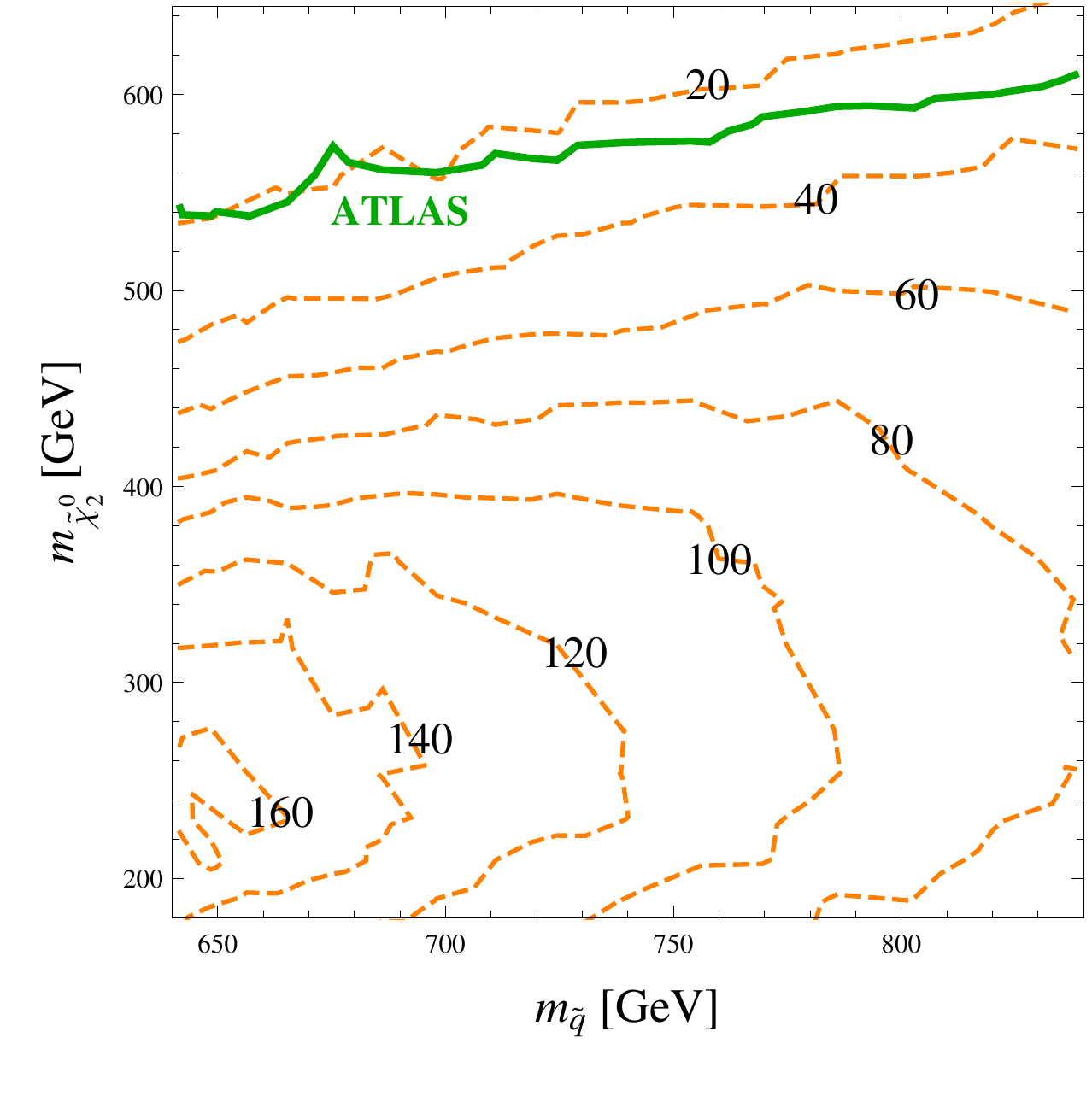}
\end{center}
\caption{ The conventional NMSSM with the same caption as Fig.~\ref{fig:scan_Di}. Noticed that
  the ATLAS+CMS limits are too stringent to be displayed in the figure.
\label{fig:scan_Ma}}
\end{figure}

It is interesting to investigate the conventional low energy NMSSM with heavy Majarona gluino,
and  compare it with our scenario, although such NMSSM cannot be realized in the
usual SUSY breaking scenarios. For this purpose, we use the same signal channels and mass spectra
as in  Dirac gluino case. The corresponding results are presented in Fig.~\ref{fig:scan_Ma}.
Form this figure, we learn the following facts. Firstly, these contours for numbers of signal events
have similar shape, but the absolute values are distinct. It is mainly due to the larger production rates
of squark productions in conventional NMSSM. Because the cut efficiencies has no huge discrepancy
for these two models, the survival event numbers are increased significantly. Secondly, for the same reason,
the exclusion limits become stringent and thus requiring more compressed spectra. To be specific,
for given $m_{\tilde{q}}\in[600,900]$ GeV, $m_{\chi^0_2}$ should heavier than $520$ GeV if one only includes
the ATLAS constraints (see green curve in Fig.~\ref{fig:scan_Ma}). Within these mass regions,
the conventional NMSSM  can also gives the best explanation for leptonic-$Z$ excess,
and the corresponding best fit benchmark point is given in Table~\ref{bench_2} as \textbf{bench3}.
However, the CMS constraints  become so stringent that almost entire parameter space in the plane
is excluded. Although a few points do survive, they contributes about $11$ signal events
in the most optimistic case (About $1.2\sigma$ signal significance,
see \textbf{bench4} in Table~\ref{bench_2}.),
and their numbers are too small to draw the exclusion curve. Therefore, only the ATLAS constraints
are displayed in Fig.~\ref{fig:scan_Ma}. The results are similar with Ref.~\cite{Cao:2015zya}, in which
the wider squark and neutralino mass ranges were studied.


\begin{table}[htbp]
\caption{The conventional NMSSM with the same caption as Table~\ref{cutflow}.}
\begin{center}
\begin{tabular}{|c||c|c|c||c|c|c|c|c|c|c|c|c||c|c|}
\hline
&$M_{\tilde{q}^{1,2}_{L,R}}$ & $M_{\tilde{\chi}^0_{1}}$ & $M_{\tilde{\chi}^0_{2}}$ & All Events& Cut1 & Cut2 & Cut3 & Cut4 & Cut5 & Cut6 & Cut7 & $N_{\rm{ Sig}}$ & $R^{\max}_{\rm{ATLAS}}$ & $R^{\max}_{\rm{ATLAS+CMS}}$  \\
\hline
\textbf{bench3}&$738$ & $489$ & $602$&$10^5$ &$14652$ &$14533$ &$13462$ &$12876$ &$2746$ &$1314$ &$1219$ & $18.5$ &$0.84$ &$1.53$  \\\hline
\textbf{bench4}&$665$ & $511$ & $624$&$10^5$ &$11941$ &$11843$ &$10959$ &$7916$ &$874$ &$398$ &$349$ &$10.6$ &$0.48$ &$0.95$  \\\hline
\end{tabular}
\label{bench_2}
\end{center}
\end{table}

\section{Conclusion}
\label{conclusion}

In the NMSSM extension with a heavy Dirac gluino and light squarks which can still keep the naturalness condition,
we interpreted the 3$\sigma$ leptonic-$Z$ excess recently reported by the ATLAS Collaboration. We concentrated on
light first two generation squark pair productions, and the corresponding decay chain is
$\tilde{q} \to q \tilde{\chi}_2^0 \to q \tilde{\chi}_1^0 Z$. Our basic strategy to produce $Z$ boson is using
a singlino-like LSP $\tilde{\chi}_1^0$ and a bino-like NLSP $\tilde{\chi}_2^0$. With specific sparticle spectra
and mass splitting $m_{\tilde{\chi}_2^0}-m_{\tilde{\chi}_1^0}\simeq 100$ GeV, the branching ratios of decays
$\tilde{q} \to q \tilde{\chi}_2^0$ and $\tilde{\chi}_2^0 \to \tilde{\chi}_1^0 Z$ are almost fixed to $100\%$
in the whole parameter space. Meanwhile, to satisfy the LHC SUSY search constraints and
maintain the nice supersafe feature of Dirac gluino,
a relatively heavy glunio is chosen, i.e., $m_{\tilde{g}}\simeq 2.6$ TeV.

Considering the bounds from all ATLAS direct SUSY searches, the leptonic-$Z$ excess can be fully addressed
in large parameter space of our model, i.e., $m_{\tilde{q}_{1,2}} \in  [650,750]$ GeV and
$m_{\tilde{\chi}^0_2} \in [200,400]$ GeV. The \textbf{bench1} with $m_{\tilde{q}_{1,2}}$=639 GeV and $m_{\tilde{\chi}^0_2}$=398 GeV
can provide the 19 signal events for the leptonic-$Z$ excess while satisfies all the
current ATLAS SUSY search constraints ($R^{\max}_{\rm{ATLAS}} = 0.56$).

The corresponding CMS search for exactly the same final states imposes very stringent bound on the above viable parameter
space which can explain the leptonic-$Z$ excess. The parameter space, which satisfies both the ATLAS and
CMS constraints, can provide at most 15 events to the excess, which means the excess can be addressed
within 1$\sigma$. The \textbf{bench2} with $m_{\tilde{q}_{1,2}}$=642 GeV and $m_{\tilde{\chi}^0_2}$=428 GeV,
which contributes 15 signal events to the excess, is on the edge of the CMS exclusion curve.

For comparison, we considered
the conventional low energy NMSSM with a heavy Majorana gluino and light squarks in the same parameter space.
The main difference is the enhanced squark pair production cross sections from  the same chirality squark pair.
In the same mass regions considered in this work, we found the capacity to explain the leptonic-$Z$ excess
in both models are similar.  If one only considered the ATLAS SUSY searches, a quite large viable parameter space
can give $\sim 19$ signal events to the $Z$ excess, e.g., \textbf{bench3}. While the constraints from
the CMS search only allow  the parameter space with less than $\sim 11$ signal events, e.g.,  \textbf{bench4}.

\section*{Acknowledgement}

This research was supported in part by the Natural Science Foundation of China under 
grant numbers 11135003, 11275246, and 11475238.


\end{document}